# Spectral shift and Q-change of circular and square-shaped optical microcavity modes due to periodic sidewall surface roughness

Svetlana V. Boriskina, Trevor M. Benson, Phillip Sewell, Alexander I. Nosich

*George Green Institute for Electromagnetics Research*
*University of Nottingham, University Park, Nottingham NG7 2RD, UK*
(June 16, 2005)

Radiation loss and resonant frequency shift due to sidewall surface roughness of circular and square high-contrast microcavities are estimated and compared by using a boundary integral equations method. An effect of various harmonic components of the contour perturbation on the Whispering-Gallery (WG) modes in the circular microdisk and WG-like modes in the square microcavity is demonstrated. In both cases, contour deformations that are matched to the mode field pattern cause the most significant frequency detuning and Q-factor change. Favorably mode-matched deformations have been found, enabling one to manipulate the Q-factors of the microcavity modes.

OCIS codes: 130.3120, 140.5960, 230.5750, 140.4780, 240.5770;   PACS numbers: 42.55.Sa, 05.45.Mt, 42.60.Da

## I. INTRODUCTION

High-index contrast optical microcavities of various shapes are versatile functional elements for dense wavelength-division multiplexing applications [1-7]. One of their most attractive features is the high quality factors theoretically achievable in very compact microcavity designs. In practice, however, the quality factors are limited by fabrication imperfections and often are much lower than their predicted values [7]. A dominant loss mechanism in semiconductor microcavities is the surface roughness [2, 8], which causes scattering of the energy of the microcavity modes and spoils their Q-factors. Thus, to realize high-Q's in microcavities it is crucial to achieve high-accuracy etching techniques to produce smooth and vertical sidewalls [8] as well as to develop analytical or numerical methods to estimate radiation loss and loss-limited Q-factors.

Surprisingly, despite the importance of modeling the effect of sidewall imperfections on the microcavity performance, few results have been published, most of them for circular microdisks operating on their WG-modes [9-12] and spherical microparticles [13]. Recently, square-shaped microcavities have attracted much attention [3-6]. Intense directional light emission observed in such cavities [6] and their potential for reducing the modal mismatch at the straight-waveguide-to-microcavity junction [4, 5] make them promising candidates as filters and laser resonators. There is still a need to study the effect of the surface roughness on the modes of square microcavities as well as discover and examine general principles of the scattering loss mechanism and frequency detuning of various types of modes in arbitrary-shape cavities.

The aim of this paper is to study how the depth and correlation length of the surface perturbations affect the natural frequencies and Q-factors of the modes in circular and square microcavities and identify the differences and common features of the radiation loss and resonant frequency detuning mechanisms. The Muller boundary integral equation (MBIE) method [14] is chosen as the simulation tool owing to its high speed, controlled accuracy, and flexibility to model arbitrary-shape smooth contours. The method demonstrates superior accuracy to the popular FDTD techniques and local-basis Galerkin algorithms. We use the effective-index approach [14] to reduce an original three-dimensional (3-D) problem of a microdisk to an equivalent 2-D formulation. In wavelength-scale thin microdisks fundamental TE-like modes (no magnetic field variation in the vertical direction; electric field in the disk plane) are dominant [1], and thus we consider only such modes in the following analysis.

## II. HIGH-Q NATURAL MODES OF CIRCULAR AND SQUARE MICRODISKS

In circular microdisks, the modes demonstrating the highest Q-factors are the Whispering-Gallery modes, with the light circulating around the rim of the cavity trapped by a quasi-total internal reflection mechanism. They are classified by azimuthal and radial mode numbers, $WG_{m,n}$, representing the number of angular and radial field variations, respectively. The modes with one radial variation of the vertical magnetic field ($WG_{m,1}$) are dominant, "true" WG-modes with the narrowest linewidths. Square optical microcavities support several types of eigenmodes [14], the most interesting for practical applications in filters and microlasers being the high-Q WG-like modes [3-6, 14] with the field nulls along the diagonals.

Using the MBIE technique, we solve the eigenvalue problems for 2-D circular and rounded-corner square microcavities having a diameter (side length) of 1.6 $\mu$m, and an effective refractive index of $n_d$=2.63. This index corresponds to the propagation constant of the TE-polarized mode at 1.55 $\mu$m in a 200 nm-thick slab of GaInAsP ($n$=3.37), which is a popular platform for semiconductor microdisk lasers [1, 15, 16]. In the vicinity of the spontaneous emission peak of the material at room temperature, 1.55-1.58 $\mu$m [16], we find a WG$_{5,1}$-mode in the circular resonator ($\lambda$=1.572 $\mu$m; Q=159.7) and a WG-like mode in the square resonator ($\lambda$=1.503 $\mu$m; Q=288.4). The mode intensity distributions (more precisely the portraits of $|H_z(x,y)|^2$) have been computed and are presented in Fig 1.

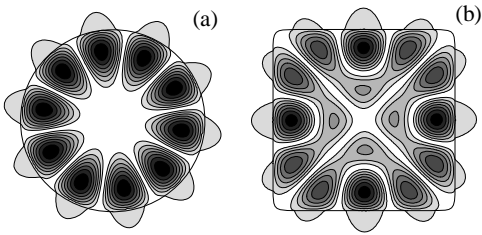

FIG. 1. Field intensity distributions for the (a) WG$_{5,1}$ mode of the circular microcavity ($\lambda$=1.572 $\mu$m) and (b) WG-like mode of the square microcavity ($\lambda$=1.503 $\mu$m). The cavities have the diameter (side length) of 1.6 $\mu$m and permittivity $\varepsilon$=6.9169+$i$10$^{-4}$.

## III. THE EFFECT OF THE CAVITY SIDEWALL ROUGHNESS

### A. Mode-matched and mismatched perturbations

To study how sidewall imperfections of various correlation lengths affect the resonant frequency and linewidths of these modes, we assume that an arbitrary contour perturbation can be decomposed into a periodic series of azimuthal harmonics [9]. We can now consider the harmonics separately and determine which of them cause the most pronounced frequency shift and Q-factor change of the microcavity modes.

We search for eigenmodes of the microcavities with corrugated boundaries described parametrically as $r(t) = R(t) + \delta\cos(\nu t)$, where $R(t)$ is a parametric expression for the undeformed cavity [14], $\delta$ is a perturbation amplitude, and parameter $t$ is the polar angle. The period (correlation length) of the perturbation normalized to the microcavity perimeter is $\Lambda = 1/\nu$. The zero perturbation harmonic corresponds to the undeformed circle (square with the corner sharpness parameter [14] equal to 10). Figs. 2 and 3 show the resonant wavelengths and Q-factors of the WG and WG-like modes as a function of the perturbation harmonic number $\nu$. Here, the characteristic perturbation amplitude is chosen as $\delta$=8 nm, based on the estimates of 5-10 nm for the mean surface roughness obtained from high-magnification SEM micrographs and scattering loss measurements [2].

WG-modes in the circular cavity are double degenerate, and as can be seen from Fig. 2, the microcavity sidewall roughness causes splitting of this degeneracy [17]. In practical realizations of circular disk microcavities, this phenomenon is manifested in double resonant peaks observed in the measurements of the laser emission spectra [1] or microdisk resonator filter response [18]. The WG-like mode in the square microcavity is a non-degenerate mode [3], and as Fig. 3 shows, the cavity boundary perturbations only shift the mode resonant frequency and change the value of its Q-factor.

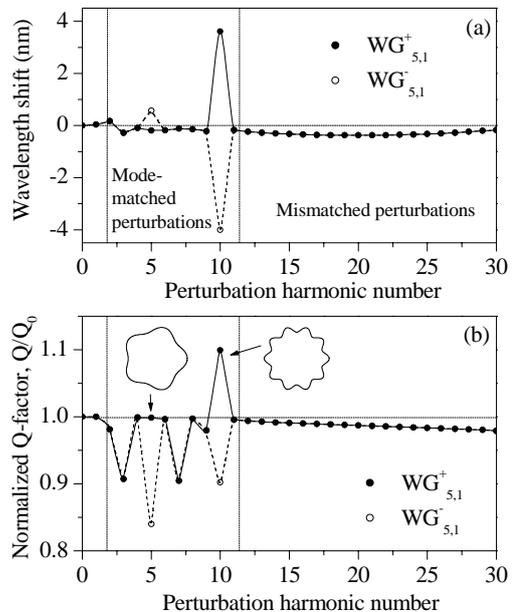

FIG. 2. Circular microcavity resonant wavelength detuning (a) and normalized loss-limited Q-factor (b) as a function of the contour perturbation harmonic number $\nu$. Radial depth of the perturbation is $\delta$=8 nm. Regions of the mode-matched perturbations and mismatched surface roughness can be observed. The inset shows two favorably mode-matched contour deformations ($\nu = 5$ and $\nu = 10$).

In both cases, there is a range of the perturbation harmonic numbers with the most noticeable variations of the cavity Q-factors and wavelengths. For the circular microcavity, this range of $\nu$ is also where the most efficient splitting of the double-degenerate WG-mode is observed. This range corresponds to the perturbation harmonics that are either favorably or unfavorably matched to the modal field distribution in the cavity. Favorably-matched corrugations increase the Q-factors of the modes, while the unfavorably-matched ones cause the degradation of the Q-factors. Mismatched contour corrugations with the values of $\nu$ ($\Lambda$) outside this range (either smaller or larger) enhance cavity scattering losses and dampen the mode Q-factors.

The width of the range of the mode-matched perturbations, as well as the behavior of the cavity Q-factor within this range, depends on the optical size of the cavity and the particular type of mode excited. For the higher-order modes this range becomes longer and shifts to higher values of $\nu$ (shorter $\Lambda$). For example, it has been shown [12] that even low-amplitude surface roughness with the correlation length $\Lambda = 0.02$ ($\nu = 50$) led to drastic degradation of the Q-factor of the $WG_{82,1}$ mode of the circular cavity.

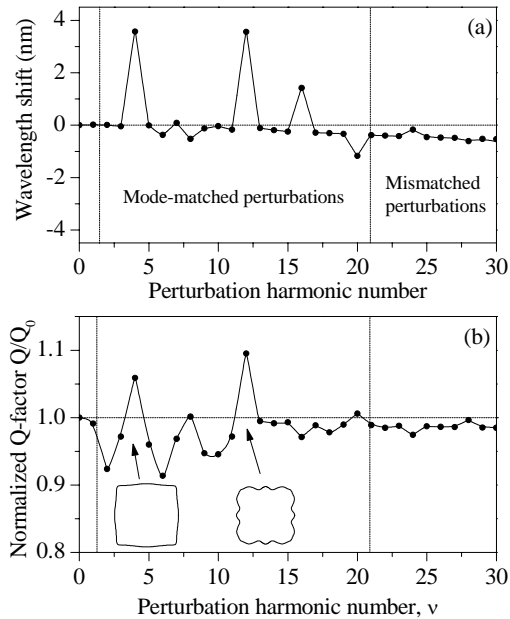

FIG. 3. Square microcavity resonant wavelength detuning (a) and normalized loss-limited Q-factor (b) as a function of the contour perturbation harmonic number $\nu$. Radial depth of the perturbation is $\delta = 8$ nm. Two types of the favorably mode-matched deformations ($\nu = 4$ and $\nu = 12$) are shown in the inset.

Furthermore, a similar grating effect has been observed when studying the influence of the surface perturbations on the Q-factors of the natural modes of 3-D spherical particles [13], though the regions of the mode-matched and mismatched perturbations have not been identified. There, certain combinations of the perturbation period and the mode number resulted in noticeable Q-spoiling of the mode, while others yielded relatively weakly Q-spoiling.

## B. Resonant frequency and Q-factor manipulation

We can observe that for the circular cavity there is one type of the most favorably mode-matched perturbation (shown in the inset of Fig. 2b) with $\nu = 2m$, which effectively splits the degenerate WG-mode, enhances one of the split modes, and suppresses the other one. This type of the corrugated microcavity has been previously proposed [15] and studied [16,17], and is called a microgear cavity. Additionally, perturbation harmonics with $\nu = m$, all other integers dividing $m$, and their products split the WG modes in more pronounced way than for other values of $\nu$ but do not enhance either of the modes.

The rapid development of modern nanotechnology calls for new microcavity shape designs that enable one to enhance the cavity lasing mode. In general, there is no *a priori* knowledge of which type of corrugation is favorably matched to a given mode type in an arbitrary-shape microcavity. Computing and plotting the values of the mode Q-factor versus the perturbation period enables us to find the deformations that enhance the cavity lasing mode. It should also be noted here that a deformation that is favorably matched to the lasing cavity mode is most likely unfavorably matched to any neighboring competing modes of different types, and thus increases the lasing mode stability and decreases its threshold.

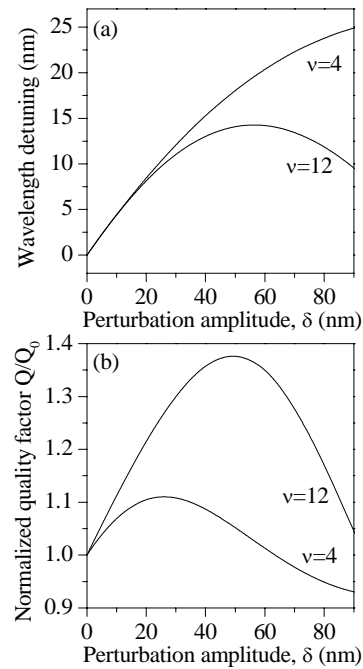

FIG. 4. Resonant wavelength shift (a) and normalized Q-factor change (b) versus the perturbation amplitude $\delta$ for the two types of the favorably matched deformations of the square microcavity ($\nu = 4$ and $\nu = 12$).

Now we apply this procedure to the square microcavity lasing on its WG-like mode, and observe (see Fig. 3b) two types of favorably mode-matched contour perturbations (shown in the inset of Fig. 3b) suitable for low-threshold semiconductor laser designs. Fig. 4 shows the resonant wavelengths (a) and Q-factors (b) of the WG-like modes in deformed square microcavities as a function of the radial depth of the perturbation. For the both types of the favorably matched perturbations an increase of the severity of the deformation results in the 11-38% increase of the mode Q-factor (Fig. 4b). As can be seen from Fig. 5, the better the cavity shape matches the mode pattern, the more efficient Q-factor manipulation can be achieved (compare to the same results for the deformed circular microcavity modes [15-17]).

It should be noted here that in some cases square cavities are formed with convex sides due to a fabrication process. E.g., the experiments with the hybrid glass square-shaped laser cavities showed that the cavity shapes became rounded because of the strong surface tension of the liquid during dip coating [6]. Our results show that such a shape deformation can be favorable for improving the microcavity performance and should not always be considered a disadvantage of the fabrication process. A similar procedure can be used to search for the deformations increasing the Q-factor of various types of modes of microcavities of any shape, e.g., triangular, racetrack, etc.

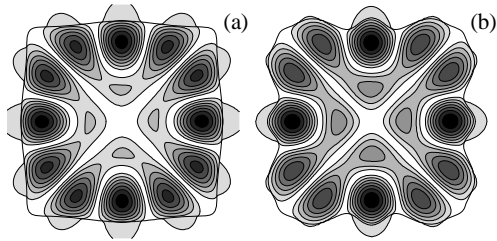

Fig. 5. Intensity patterns of the WG-like modes in the deformed square microcavities. The radial depth of each perturbation corresponds to the points of the maxima of the mode Q-factor in Fig. 4b ($\delta_{\nu=4}$=28 nm, $\delta_{\nu=12}$=48 nm).

### C. The effect of the low-amplitude mismatched roughness

Finally, we investigate the effect of the amplitude $\delta$ of the mismatched perturbation ($\Lambda = 0.033$) on the cavity characteristics and observe a blueshift of the lasing mode frequencies and degradation of their Q-factors if $\delta$ is increased (Fig. 6 a,b). Frequency shift in both cases (Fig. 6 a) is caused by a shrinkage of the effective cavity size due to the increased amplitude of the contour perturbations. At the same time, growing scattering losses (Fig. 6 b) dampen the Q-factors. However, we can see that the cavity characteristics are almost insensitive to low-amplitude ($\delta < 8-10$ nm) perturbations. The same effect has been observed for mismatched perturbation harmonics of different correlation lengths ($\Lambda = 0.0357$ and $\Lambda = 0.0313$) and thus reflects a common feature of microcavity modes.

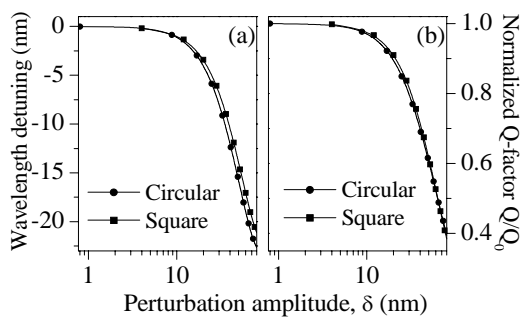

FIG. 6. Resonant wavelength detuning (a) and Q-factor degradation (b) of the $WG_{5,1}$-mode of the circular microdisk and WG-like mode of the square microcavity as a function of the depth of the mismatched contour perturbation with the period $\Lambda$=0.033 ($\nu$=30).

Furthermore, the same threshold-like behavior of the Q-factor as a function of the perturbation amplitude has been observed in the spherical cavity, for both, the periodical and irregular surface roughness (see Ref [13], Figs. 2 and 4). There, the presence of the surface perturbations had minimal effect on the resonance Q-factor until reaching a certain threshold value of the perturbation amplitude, beyond which the Q-factor degraded rapidly. This feature of the natural modes of dielectric cavities makes possible the experimentally achieved high-Q oscillations, even though every fabricated microcavity is slightly imperfect [1, 2, 6, 7].

### IV. CONCLUSIONS

In summary, we have shown that the performance of high-index contrast microcavities is strongly affected by surface-roughness-induced radiation loss, this effect being most pronounced for roughness distributions that are matched to the modal field pattern in the cavity. A possibility for mode Q-factor manipulation by controlling special mode-matched boundary perturbations is demonstrated. The contour perturbation with the period $\Lambda = (2m)^{-1}$, yielding a microgear cavity [15-17], is shown to be the best mode-matched deformation of the circular cavity that enables one to effectively split the double-degenerate $WG_{m,1}$-mode and achieve Q-factor control. By varying the shape and depth of the boundary perturbations of the square microcavity we demonstrated an increase of the cavity Q-factor by up to 38%. Stability of the mode characteristics against low-amplitude perturbations is shown, enabling us to estimate microcavity fabrication tolerances. Furthermore, our comparison with published results proves that the principles demonstrated here for ultrasmall 2-D cavities are directly applicable to larger 2-D [12] and 3-D [13] dielectric resonators.

### Acknowledgements

This work has been supported by the UK Engineering and Physical Sciences Research Council (EPSRC) under Grants GR/R65213/01 and GR/S60693/01(P). S. Boriskina's e-mail address is eezsb@gwmail.nottingham.ac.uk.